# 車聯網安全憑證管理系統效能分析

| Abel C. H. Chen | Cheng-Kang Liu | Chun-Feng Lin | Bon-Yeh Lin |
| --- | --- | --- | --- |
| 中華電信研究院 | 中華電信研究院 | 中華電信研究院 | 中華電信研究院 |
| 資通安全研究所 | 資通安全研究所 | 資通安全研究所 | 資通安全研究所 |
| 高級研究員 | 高級研究員 | 高級研究員 | 資深研究員 |
| 0000-0003-3628-3033 | toniliu@cht.com.tw | landbird@cht.com.tw | bylin@cht.com.tw |

## 摘要

在車聯網(Vehicle-to-Everything, V2X)通訊過程中,提供正確資訊和保護終端設備隱私是重要的資訊安全議題之一。有鑑於此,近幾年國際標準制定組織開始制定車聯網通訊安全標準;例如:IEEE 1609.2.1 標準中設計安全憑證管理系統(Security Credential Management System, SCMS),規範憑證申請和簽發流程,以及憑證廢止流程等。並且在 IEEE 1609.2 標準中定義憑證格式和安全協定資料單元(Secure Protocol Data Unit, SPDU),讓資料可以建構在安全標準上進行傳輸。因此,世界各國的終端設備商和安全憑證提供商開始根據標準建構車聯網安全系統,並且進行互通測試。雖然在國際標準中主要採用橢圓曲線密碼學(Elliptic-Curve Cryptography, ECC)來提供簽章/驗章和加密/解密功能,但效能分析是系統實際上路的重要關鍵議題之一。因此,本研究實作符合 IEEE 1609.2 標準、IEEE 1609.2.1 標準的終端設備和安全憑證管理系統,並且從終端設備的視角來測量在系統中每個安全通訊行為的計算和傳輸時間,並且找出潛在的系統瓶頸。在實驗結果中,本研究分析最耗費系統效能的行為,並提出相關系統效能提升建議供安全憑證管理系統開發者參考。

關鍵字:車聯網、安全憑證管理系統、效能分析。

## 一、前言

近年來,隨著車聯網(Vehicle-to-Everything, V2X)通訊技術的成熟,各式各樣的智慧型運輸系統應用也開始蓬勃發展。隨之而來的則是開始注重車聯網通訊的安全性和終端設備的隱私,建立安全通訊環境。因此,已經有多個國家/地區(如:美國)開始建立試驗區,實施符合 IEEE 1609.2 標準和 IEEE 1609.2.1 標準的安全憑證管理系統(Security Credential Management System, SCMS) [1]-[2]。終端設備(End Entities, EEs)可以跟安全憑證管理系統申請憑證,並且對其發送的安全協定資料單元(Secure Protocol Data Unit, SPDU)進行簽章,以及對其接收的安全協定資料單元進行驗章,建立安全可信的車聯網通訊。

在 IEEE 1609.2 標準和 IEEE 1609.2.1 標準中主要採用橢圓曲線密碼學(Elliptic-Curve Cryptography, ECC) [3]-[4]作為非對稱金鑰密碼學方法,並且採用橢圓曲線數位簽章演算法(Elliptic Curve Digital Signature Algorithm, ECDSA) [5]進行簽章和驗章、採用橢圓曲線整合加密方案(Elliptic Curve Integrated Encryption Scheme, ECIES) [6]-[7]進行加密和解密。每張憑證和每個安全協定資料單元可以用橢圓曲線數位簽章演算法進行簽章和驗章,決定憑證和安全協定資料單元的內容是否可信。此外,IEEE 1609.2.1 標準中提出蝴蝶金鑰擴展(Butterfly Key Expansion, BKE)機制,可以提供假名憑證(pseudonymous certificate)給終端設備,保護終端設備的隱私。

IEEE 1609.2.1 標準也規範了申請和簽發註冊憑證(Enrollment Certificate, EC)和授權憑證(Authorization Certificate, AC)的流程及其封包結構[1]-[2]。有鑑於此,本研究根據 IEEE 1609.2 標準和 IEEE 1609.2.1 標準實作安全憑證管理系統,並且量測每個安全通訊

行為的計算和傳輸時間，進行系統效能分析和探討瓶頸，以及在第六節中提出相關建議改善系統效能。

本文分為七個部分。第二節介紹安全憑證管理系統的架構，並且描述每個憑證機構的角色。第三節演示蝴蝶金鑰擴展機制的流程，從建立毛蟲金鑰對到取得蝴蝶金鑰對的步驟。第四節對應第三節的蝴蝶金鑰擴展機制，說明從申請註冊憑證到取得授權憑證(即假名憑證)的流程。第五節概述安全協定資料單元的結構，並且以車載設備(On-Board Unit, OBU)發送基本安全訊息(Basic Safety Message, BSM)和路側設備(Road-Side Unit, RSU)發送號誌時相(Signal Phase and Timing, SPaT)訊息為例。第六節討論安全憑證管理系統的詳細性能分析和建議。最後，第七節總結本研究的貢獻和討論未來研究。

## 二、安全憑證管理系統架構

本研究在圖一展示簡化版安全憑證管理系統，包含根憑證中心(Root Certificate Authority, RCA)、中繼憑證中心(Intermediate Certificate Authority, ICA)、授權憑證中心(Authorization Certificate Authority, ACA)/假名憑證中心(Pseudonym Certificate Authority, PCA)、註冊憑證中心(Enrollment Certificate Authority, ECA)、登錄中心(Registration Authority, RA)和終端設備[2]。其中，在 IEEE 1609.2.1 標準的 2022 年版將假名憑證中心更名為授權憑證中心，假名憑證更名為授權憑證。

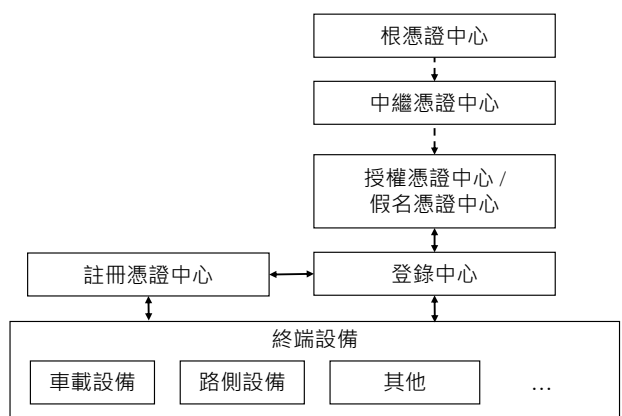

圖一：簡化版安全憑證管理系統[2]

在安全憑證管理系統中，選舉人(Elector)可以在憑證信任列表(Certificate Trust List, CTL)中簽署可信任的根憑證中心，並且其他設備將以憑證信任列表作為信任錨點(Trust Anchor)。根憑證中心可以簽發中繼憑證中心的憑證，並且中繼憑證中心可以簽發授權憑證中心/假名憑證中心、註冊憑證中心、登錄中心的憑證。

終端設備(包含車載設備和路側設備)需在出廠前先在註冊憑證中心註冊權威公鑰(Canonical Public Key)，在開機時用權威私鑰(Canonical Private Key)簽署註冊憑證請求，在請求內容帶入註冊公鑰(Enrollment Public Key)，向註冊憑證中心申請取得註冊憑證。取得註冊憑證後，再以註冊私鑰(Enrollment Private Key)簽署授權憑證請求，向登錄中心申請取得授權憑證[2]。註冊憑證和授權憑證的具體流程將在第三節和第四節中介紹。

## 三、蝴蝶金鑰擴展機制

IEEE 1609.2.1 標準中提出蝴蝶金鑰擴展機制，用來產製假名憑證，以提高車聯網通訊中車載設備的隱私，如圖二所示。其中，蝴蝶金鑰擴展機制的步驟包括產製毛蟲金鑰對(Caterpillar Key Pairs)、產製繭公鑰(Cocoon Public Keys)、產製蝴蝶公鑰(Butterfly

Public Key)，以及產製繭私鑰(Cocoon Private Keys)和蝴蝶私鑰(Butterfly Private Key)[2]。

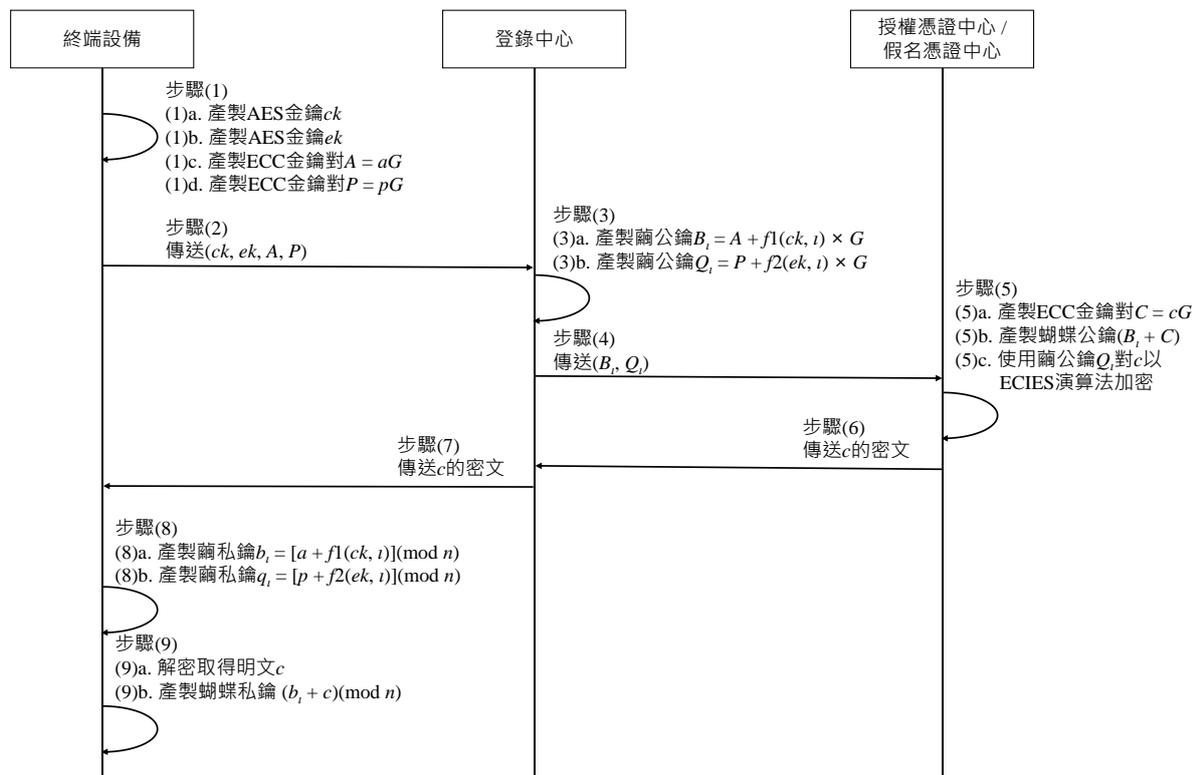

圖二：蝴蝶金鑰擴展機制[2]

### 3.1 產製毛蟲金鑰對
- 步驟(1)a：產製進階加密標準(Advanced Encryption Standard, AES)金鑰 $ck$ 作為簽章/驗章用。
- 步驟(1)b：產製 AES 金鑰 $ek$ 作為加密用。
- 步驟(1)a：產製 ECC 金鑰對 $A = aG$ 作為簽章/驗章用毛蟲金鑰對。其中，$a$ 為私鑰(如第二節提及的註冊私鑰)、$A$ 為公鑰(如第二節提及的註冊公鑰)、$G$ 為橢圓曲線的基點座標。
- 步驟(1)a：產製 ECC 金鑰對 $P = pG$ 作為加密/解密用毛蟲金鑰對。其中，產生隨機數 $p$ 作為私鑰，並且對應的公鑰為 $P$。
- 步驟(2)：向登錄中心申請授權憑證，在授權憑證請求中傳送($ck, ek, A, P$)給登錄中心。

### 3.2 產製繭公鑰
- 步驟(3)a：產製繭公鑰 $B_\iota = A + f1(ck, \iota) \times G$ 作為簽章/驗章用繭公鑰。其中，$f1(ck, \iota)$ 為擴展函數，根據時間週期和索引值產生 $\iota$，再以 AES 金鑰 $ck$ 加密 $\iota$ 後產生擴展值，在毛蟲公鑰 $A$ 基礎上擴展得到繭公鑰 $B_\iota$。
- 步驟(3)b：產製繭公鑰 $Q_\iota = P + f2(ek, \iota) \times G$ 作為加密/解密用繭公鑰。其中，$f2(ek, \iota)$ 為擴展函數，根據時間週期和索引值產生 $\iota$，再以 AES 金鑰 $ek$ 加密 $\iota$ 後產生擴展值，在毛蟲公鑰 $P$ 基礎上擴展得到繭公鑰 $Q_\iota$。
- 步驟(4)：向授權憑證中心申請授權憑證，在授權憑證請求中傳送($B_\iota, Q_\iota$)給授權憑證中心。

### 3.3. 產製蝴蝶公鑰

- 步驟(5)a：產製 ECC 金鑰對 $C = cG$。其中，產生隨機數 $c$ 作為私鑰，並且對應的公鑰為 $C$，讓每次產製的蝴蝶金鑰對可以有不同的值，以提升安全性和保護終端設備隱私。
- 步驟(5)b：產製蝴蝶公鑰($B_i + C$)，並且存放在授權憑證中，後續可用此公鑰驗證對應的私鑰簽發的安全協定資料單元。
- 步驟(5)c：使用繭公鑰 $Q_i$ 對 $c$ 以 ECIES 演算法加密，避免傳送過程中被登錄中心取得 $c$，用以保護終端設備隱私。
- 步驟(6)：由授權憑證中心回覆登錄中心授權憑證請求結果，並且內容帶有 $c$ 的密文給登錄中心。
- 步驟(7)：由登錄中心回覆終端設備授權憑證請求結果，並且內容帶有 $c$ 的密文給終端設備。

### 3.4 產製繭私鑰和蝴蝶私鑰

- 步驟(8)a：產製繭私鑰 $b_i = [a + f1(ck, i)] \pmod{n}$ 作為簽章/驗章用繭私鑰。根據擴展函數 $f1(ck, i)$、時間週期和索引值 $i$、以及以 AES 金鑰 $ck$ 加密 $i$ 後對毛蟲私鑰 $a$ 進行擴展，得到擴展後的繭私鑰 $b_i$。
- 步驟(8)a：產製繭私鑰 $q_i = [p + f2(ek, i)] \pmod{n}$ 作為加密/解密用繭私鑰。根據擴展函數 $f2(ek, i)$、時間週期和索引值 $i$、以及以 AES 金鑰 $ek$ 加密 $i$ 後對毛蟲私鑰 $p$ 進行擴展，得到擴展後的繭私鑰 $q_i$。
- 步驟(9)a：運用繭私鑰 $q_i$ 對 $c$ 的密文進行解密，取得明文 $c$。
- 步驟(9)b：產製蝴蝶私鑰 $(b_i + c) \pmod{n}$。以明文 $c$ 對繭私鑰 $b_i$ 進行擴展，得到擴展後的蝴蝶私鑰。

## 四、憑證請求流程

本節將在 4.1 節介紹申請和取得註冊憑證的流程，以及 4.2 節介紹申請和取得授權憑證(即假名憑證)的流程，如圖三所示[2]。

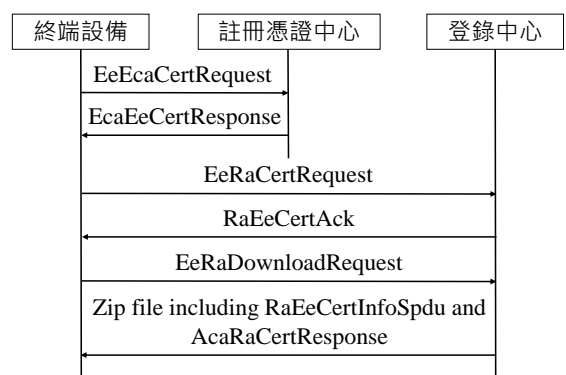

圖三：憑證請求流程[2]

### 4.1 註冊憑證請求流程

- **初始階段**：終端設備(包含車載設備和路側設備)需在出廠前先在註冊憑證中心註冊權威公鑰(Canonical Public Key)，並且終端設備持有權威私鑰(Canonical Private Key)。
- **EeEcaCertRequest**：由終端設備發送 EeEcaCertRequest 請求給註冊憑證中心，

向註冊憑證中心申請註冊憑證。其中，終端設備隨機產生註冊金鑰對(Enrollment Key Pair)，包含註冊私鑰(Enrollment Private Key)(即蝴蝶金鑰擴展機制中的毛蟲私鑰 $a$)和註冊公鑰(Enrollment Public Key)(即蝴蝶金鑰擴展機制中的毛蟲公鑰 $A$)；終端設備將註冊公鑰設定到 EeEcaCertRequest 請求中，並且採用權威私鑰對 EeEcaCertRequest 請求簽章，證實確實是由該合法終端設備發送的請求。

- **EcaEeCertResponse**：由註冊憑證中心簽發註冊憑證，並發送 EcaEeCertResponse 回應給終端設備。其中，由註冊憑證中心收到 EeEcaCertRequest 請求後，根據已註冊的權威公鑰(Canonical Public Key)進行驗章，確認確實為該合法終端設備發送的請求後，再根據 EeEcaCertRequest 請求簽發註冊憑證，並且該註冊憑證帶有該終端設備的註冊公鑰。之後再把註冊憑證放到 EcaEeCertResponse 回應中，並且發送 EcaEeCertResponse 回應給該終端設備。

## 4.2 授權憑證請求流程

- **EeRaCertRequest**：由終端設備發送 EeRaCertRequest 請求給登錄中心，向登錄中心申請授權憑證。其中，終端設備隨機產生一組加密/解密用毛蟲金鑰對(即蝴蝶金鑰擴展機制中的 $P = pG$)、一組金鑰 $ck$ 作為簽章/驗章用、一組 AES 金鑰 $ek$ 作為加密用，連同帶有註冊公鑰的註冊憑證及相關請求內容(如：提供者服務識別碼(Provider Service Identifier, PSID)等資訊)設定到 EeRaCertRequest 請求，並且發送給登錄中心。
- **RaEeCertAck**：由登錄中心確認終端設備合法性及其請求內容，並且向授權憑證中心請求簽發授權憑證，並發送 RaEeCertAck 回應給終端設備。其中，由註冊憑證中心收到 EeRaCertRequest 請求後，取得註冊憑證中心憑證和公鑰驗證終端設備註冊憑證的合法性，確認確實為該合法終端設備發送的請求後，再根據 RaEeCertAck 請求向授權憑證中心請求簽發授權憑證。過程中將根據根據時間週期和索引值產生 $i$，搭配擴展函數、AES 金鑰、以及毛蟲公鑰產製多把繭公鑰，並且向授權憑證中心發出授權憑證請求。之後再把受理結果和 EeRaCertRequest 請求的雜湊值(hash)放到 RaEeCertAck 回應中，並且發送 RaEeCertAck 回應給該終端設備。
- **EeRaDownloadRequest**：由終端設備發送 EeRaDownloadRequest 請求給登錄中心，向登錄中心申請下載授權憑證。其中，終端設備在 EeRaDownloadRequest 請求設定 EeRaCertRequest 請求的雜湊值，讓登錄中心可根據 EeRaCertRequest 請求的雜湊值提供對應的授權憑證。
- Zip 檔：由登錄中心根據 EeRaCertRequest 請求的雜湊值，將授權憑證中心簽發授權憑證及蝴蝶金鑰擴展機制中的每張授權憑證中 $c$ 的密文封裝縮到 Zip 檔，並且提供給終端設備。終端設備下載 Zip 檔後，解壓縮後可以取得 RaEeCertInfoSpdu 和 AcaRaCertResponse，並且可以在 AcaRaCertResponse 取得權憑證及蝴蝶金鑰擴展機制中的每張授權憑證中 $c$ 的密文。終端設備可以根據 AcaRaCertResponse 計算取得繭私鑰、蝴蝶私鑰、以及授權憑證。後續終端設備可以將該授權憑證設定到各種智慧型運輸系統訊息類型的安全協定資料單元，並且採用蝴蝶私鑰對安全協定資料單元(具體結構在第五節中描述)進行簽章，而其他設備可以採用授權憑證中的蝴蝶公鑰進行驗章。

## 五、安全協定資料單元結構

IEEE 1609.2 標準定義的安全協定資料單元結構主要包含簽署資料(Signed Data)，而在簽署資料中包含被簽署資料(To-Be-Signed Data, TBSD)、簽章者識別碼(Signer Identifier, SI)、以及簽章(Signature)，如圖四所示。其中，被簽署資料包含有標頭資訊(Header Info, HI)、簽署資料內容(Signed Data Payload)，並且智慧型運輸系統的應用服務訊息將放置在簽署資料內容中傳送。簽章者識別碼可以用來取得簽章者資訊，如：簽章者的憑證或憑證摘要(Digest)；而簽章欄位主要存放簽章者用其私鑰對被簽署資料的簽章資訊，通過簽章者識別碼和簽章可以用來驗證該安全協定資料單元是否為可信的資料。在 5.1 節中，以車載設備常發送的基本安全訊息為例，說明基於基本安全訊息的安全協定資料單元；在 5.2 節中，以路側設備常發送的號誌時相為例，說明基於號誌時相的安全協定資料單元。

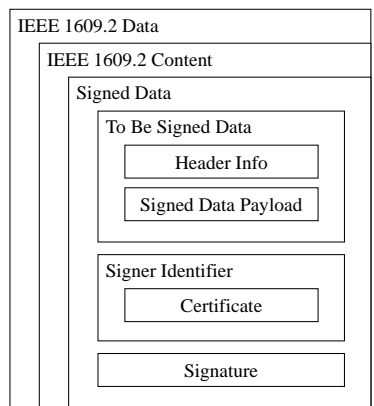

圖四：安全協定資料單元結構[1]

### 5.1 基於基本安全訊息的安全協定資料單元

汽車工程師學會(Society of Automotive Engineers, SAE)已經針對基本安全訊息制定相關格式標準，並且定義在 SAE J2735 標準文件[8]。例如：圖五顯示基本安全訊息的 JavaScript 目標表示(JavaScript Object Notation, JSON)格式的例子，並且可對該基本安全訊息 JSON 格式的內容以十六進位(Hexadecimal, Hex)編碼為"0014334d185898d939082242691859c0f78f080000000000000180007e7d07d07f7fff17e0000001001d640020f9d902001034208820"。車載設備可將該十六進位編碼作為簽署資料內容，並且在標頭資訊加入產製時間(Generation Time)，以及用車載設備的蝴蝶私鑰進行簽章產生基於基本安全訊息的安全協定資料單元。

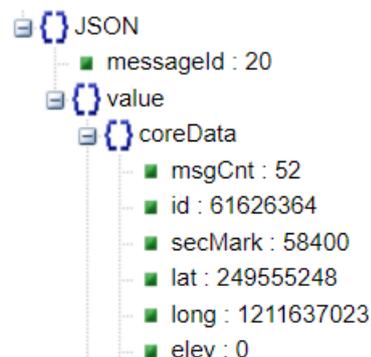

圖五：基本安全訊息 JSON 格式案例

## 5.2 基於號誌時相的安全協定資料單元

汽車工程師學會已經針對號誌時相制定相關格式標準，並且定義在 SAE J2735 標準文件[8]。例如：圖六顯示號誌時相的 JSON 格式的例子，並且可對該基本安全訊息 JSON 格式的內容以十六進位編碼為"001314000007d48904000100004303dc300082401ee180"。路側設備可將該十六進位編碼作為簽署資料內容，並且在標頭資訊除了加入產製時間外，還需要加入產製位置(Generation Location)，以及用路側設備的蝴蝶私鑰進行簽章產生基於號誌時相的安全協定資料單元。

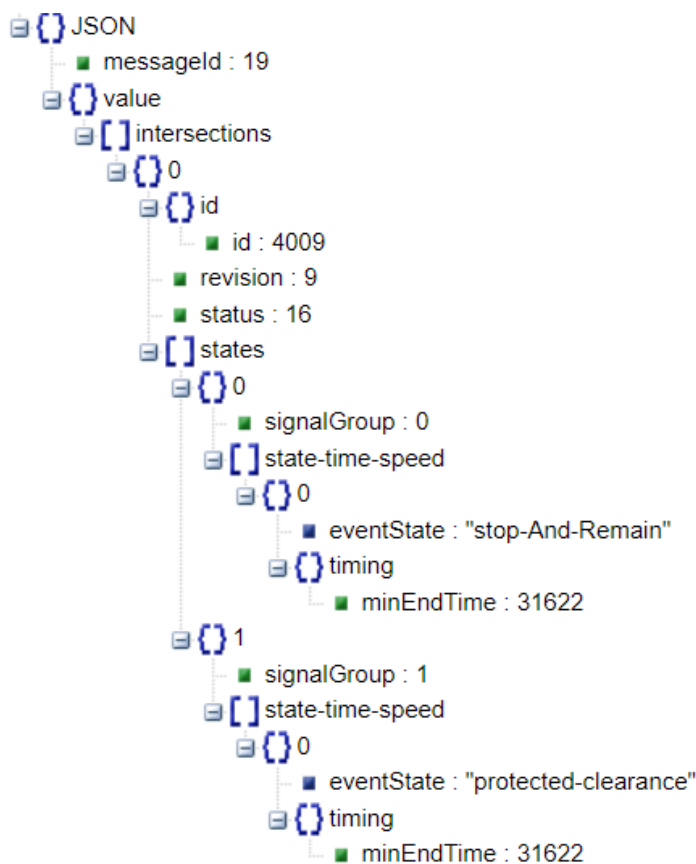

圖六：號誌時相 JSON 格式案例

## 六、實驗環境與結果討論

本節將在 6.1 節介紹實驗環境，並且定義測試目標。在 6.2 節介紹實驗結果，並且在 6.3 節討論系統效能提升建議。

### 6.1 實驗環境

在實驗環境中，本研究選擇中華電信作為安全憑證管理系統提供商進行效能驗證。其中，中華電信作為正式的安全憑證管理系統提供商資訊可在 OmniAir 748 [9]文件中取得，例如：註冊憑證中心的憑證識別碼為"eca.cht-scms.com.tw"。終端設備可以先把其權威公鑰註冊在註冊憑證中心，後續再發送 EeEcaCertRequest 請求，向註冊憑證中心取得註冊憑證。並且，終端設備可以執行圖三後續步驟，向登錄中心取得授權憑證，以及發送安全協定資料單元。

為實證系統效能，本研究採用公信電子(Clientron)的設備作為終端設備，並且開發終端設備 App 植入到設備中執行，如圖七所示。該設備的作業系統是 Android 10，並且 App 採用 Java 語言開發。在實驗環境中，本研究開發一台車載設備、一台路側設備，

分別發送基於基本安全訊息的安全協定資料單元和基於基本安全訊息的安全協定資料單元。

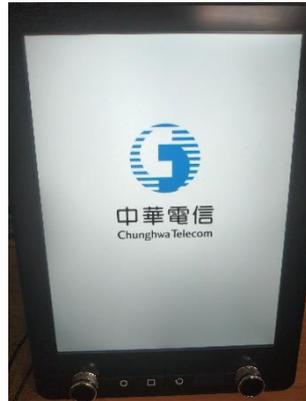

圖七：中華電信終端設備 App 安裝於公信電子設備

為驗證安全憑證管理系統的系統效能，本研究將過程拆解成 13 種行為，並分別對其進行測量。

- **Action 1：編碼和簽署 EeEcaCertRequest**。此行為由終端設備將隨機產生註冊金鑰對，並且將註冊公鑰設定到 EeEcaCertRequest 請求中，並且對該請求運用權威私鑰簽章。
- **Action 2：發送 EeEcaCertRequest 和接收 EcaEeCertResponse**。此行為由終端設備發送 EeEcaCertRequest 給註冊憑證中心，再從註冊憑證中心接收 EcaEeCertResponse。過程中，主要是由註冊憑證中心對 EeEcaCertRequest 請求進行驗章，並產製註冊憑證，以及對註冊憑證和 EcaEeCertResponse 簽章。
- **Action 3：解碼和驗證 EcaEeCertResponse**。此行為由終端設備在取得 EcaEeCertResponse 後進行解碼，並且驗證 EcaEeCertResponse 裡面的簽章。
- **Action 4：編碼和簽署 EeRaCertRequest**。此行為由終端設備將隨機產生一組 ECC 金鑰對和兩組 AES 金鑰，並且將金鑰和註冊憑證設定到 EeRaCertRequest 請求中，並且對該請求運用註冊私鑰簽章。
- **Action 5：發送 EeRaCertRequest 和接收 RaEeCertAck**。此行為由終端設備發送 EeRaCertRequest 給登錄中心，再從登錄中心接收 RaEeCertAck。過程中，主要是由登錄中心對 EeRaCertRequest 請求進行驗章，通過驗證後產製繭公鑰，並向授權憑證中申請授權憑證，對 RaEeCertAck 編碼和簽章後發送給終端設備。
- **Action 6：解碼和驗證 RaEeCertAck**。此行為由終端設備在取得 RaEeCertAck 後進行解碼，並且驗證 RaEeCertAck 裡面的簽章。
- **Action 7：編碼和簽署 EeRaDownloadRequest**。此行為由終端設備將 EeRaCertRequest 雜湊值設定到 EeRaDownloadRequest 請求中，並且對該請求運用註冊私鑰簽章。
- **Action 8：發送 EeRaDownloadRequest 和接收 Zip 檔**。此行為由終端設備發送 EeRaDownloadRequest 給登錄中心，再從登錄中心接收 Zip 檔。過程中，主要是由登錄中心對 EeRaDownloadRequest 請求進行驗章，封裝 RaEeCertInfoSpdu 和 AcaRaCertResponse 到 Zip 檔，並且發送 Zip 檔給終端設備。
- **Action 9：解碼和驗證 Zip 檔**。此行為由終端設備在取得 Zip 檔後進行解碼，

取得 RaEeCertInfoSpdu 和 AcaRaCertResponse，並且驗證裡面的簽章。驗證簽章無誤後，再根據每一包 AcaRaCertResponse 的內容取得授權憑證，以及產製蝴蝶私鑰。

- **Action 10**：**編碼和簽署具有完整憑證的安全協定資料單元**。此行為由車載設備設定基本安全訊息到安全協定資料單元、由路側設備設定號誌時相到安全協定資料單元，並且各別將完整授權憑證設定到安全協定資料單元的簽章者識別碼，以及使用其蝴蝶私鑰對安全協定資料單元的被簽署資料進行簽章。
- **Action 11**：**編碼和簽署具有憑證摘要的安全協定資料單元**。此行為由車載設備設定基本安全訊息到安全協定資料單元、由路側設備設定號誌時相到安全協定資料單元，並且各別將憑證摘要設定到安全協定資料單元的簽章者識別碼，以及使用其蝴蝶私鑰對安全協定資料單元的被簽署資料進行簽章。
- **Action 12**：**解碼和驗證具有完整憑證的安全協定資料單元**。此行為由終端設備接收安全協定資料單元，並且根據簽章者識別碼取得對應的授權憑證和產製該簽章者的蝴蝶公鑰，以及使用蝴蝶公鑰進行驗章。
- **Action 13**：**解碼和驗證具有憑證摘要的安全協定資料單元**。此行為由終端設備接收安全協定資料單元，並且根據簽章者識別碼取得對應的授權憑證和產製該簽章者的蝴蝶公鑰，以及使用蝴蝶公鑰進行驗章。

## 6.2 實驗結果

為驗證安全憑證管理系統的系統效能，本研究對 6.1 節定義的 13 種行為各執行 20 次，實驗結果整理於表一。其中，為保護車載設備隱私，所以車載設備每次申請授權憑證時，將一次簽發 20 張授權憑證(即假名憑證)，讓車載設備可以採用不同的授權憑證簽發安全協定資料單元。在路側設備則不具隱私議題，所以路側設備每次申請授權憑證時，將一次簽發 1 張授權憑證，讓路側設備採用同一張的授權憑證簽發安全協定資料單元。

表一：各種行為的平均時間長度(單位：毫秒)

| 行為 | OBU | RSU |
| --- | --- | --- |
| **Action 1**：編碼和簽署 EeEcaCertRequest | 46 | 45 |
| **Action 2**：發送 EeEcaCertRequest 和接收 EcaEeCertResponse | 322 | 259 |
| **Action 3**：解碼和驗證 EcaEeCertResponse | 396 | 431 |
| **Action 4**：編碼和簽署 EeRaCertRequest | 161 | 148 |
| **Action 5**：發送 EeRaCertRequest 和接收 RaEeCertAck | 1249 | 281 |
| **Action 6**：解碼和驗證 RaEeCertAck | 92 | 79 |
| **Action 7**：編碼和簽署 EeRaDownloadRequest | 140 | 141 |
| **Action 8**：發送 EeRaDownloadRequest 和接收 Zip 檔 | 1144 | 104 |
| **Action 9**：解碼和驗證 Zip 檔 | 13026 | 348 |
| **Action 10**：編碼和簽署具有完整憑證的安全協定資料單元 | 26 | 28 |
| **Action 11**：編碼和簽署具有憑證摘要的安全協定資料單元 | 23 | 25 |
| **Action 12**：解碼和驗證具有完整憑證的安全協定資料單元 | 125 | 130 |
| **Action 13**：解碼和驗證具有憑證摘要的安全協定資料單元 | 124 | 123 |

由實驗結果可得知，花費最多時間的行為是 Action 9 (解碼和驗證 Zip 檔)，車載設備需花費 13026 毫秒；有鑑於車載設備一次需驗證 20 包 AcaRaCertResponse 的簽章及

產製 20 組繭私鑰和蝴蝶私鑰，所以花費大量的計算時間。除此之外，Action 5 (發送 EeRaCertRequest 和接收 RaEeCertAck)和 Action 8 (發送 EeRaDownloadRequest 和接收 Zip 檔)也是需要花費較多的時間，車載設備需各 1249 毫秒和 1144 毫秒；其原因主要在於需一次產製和下載 20 張授權憑證，然而大量計算工作在於登錄中心和授權憑證中心，所以對系統較不會造成瓶頸，主要時間花費在傳輸上。

由於 IEEE 1609.2 標準和 IEEE 1609.2.1 標準主要採用橢圓曲線密碼學，在橢圓曲線密碼學建立簽章和驗章，所以橢圓曲線數位簽章演算法為主要採用的方法。然而，橢圓曲線數位簽章演算法在簽章時效率較高，驗章時需要較多的計算時間[10]。因此，從實驗結果可以看到 Action 3 (解碼和驗證 EcaEeCertResponse)的計算時間比 Action 1 (編碼和簽署 EeEcaCertRequest)的計算時間長。相同的，Action 12 (解碼和驗證具有完整憑證的安全協定資料單元)的計算時間比 Action 10 (編碼和簽署具有完整憑證的安全協定資料單元)的計算時間長；Action 13 (解碼和驗證具有憑證摘要的安全協定資料單元)的計算時間比 Action 11 (編碼和簽署具有憑證摘要的安全協定資料單元)的計算時間長。

### 6.3 討論與建議

有鑑於蝴蝶金鑰擴展機制所需花費的計算時間與產製新的金鑰對的計算時間相似，需要較多的計算時間，如：表一中的 Action 9 (解碼和驗證 Zip 檔)。為提升安全憑證管理系統的系統效能，本研究整理下面幾點建議：

- 將已經產製的蝴蝶金鑰暫存在安全晶片中，避免每次使用時動態擴展蝴蝶金鑰，耗費計算時間。
- 由於執行蝴蝶金鑰擴展機制的時間點是可預期的，所以可以在離峰時進行。
- 可以在背景程式批次擴展蝴蝶金鑰，避免佔用系統資源。

## 七、結論與未來研究

本研究實作符合 IEEE 1609.2 標準和 IEEE 1609.2.1 標準的安全憑證管理系統，並且從終端設備的角度來評估安全憑證管理系統的系統效能。在驗證過程中，從發送 EeEcaCertRequest 請求到取得 RaEeCertInfoSpdu 和 AcaRaCertResponse 的 Zip 檔，驗證編碼、解碼、簽章、驗章，以及組建蝴蝶金鑰的計算時間等。最後，採用車載設備和路側設備分別收發基於基本安全訊息的安全協定資料單元和基於號誌時相的安全協定資料單元，驗證每個行為所需的計算時間，並且找出系統瓶頸。根據找到的系統瓶頸，本研究提出效能提升建議，供後續安全憑證管理系統開發者參考。

在未來研究中，可以測量和分析更多的行為(例如：憑證信任列表(Certificate Trust List, CTL)下載和憑證註銷列表(Certificate Revocation List, CRL)下載)，以提升安全憑證管理系統的系統效能。此外，未來可研究後量子密碼方法，並且結合到安全憑證管理系統，增加抗量子計算的能力。

## 參考文獻